\documentclass[useapjfonts,apj]{emulateapj}
\usepackage{psfrag,color}
\usepackage{commath}
\usepackage{amssymb}

\def\mathnew{\mathsurround=0pt}
\def\simov#1#2{\lower .5pt\vbox{\baselineskip0pt
    \lineskip-.5pt\ialign{$\mathnew#1\hfil##\hfil$\crcr#2\crcr\sim\crcr}}}  
\def\simgreat{\mathrel{\mathpalette\simov >}}
\def\simless{\mathrel{\mathpalette\simov <}}

\def\alamenos#1{$^{-#1}$}
\def\cmtres{\mbox{cm$^{-3}$}}

\def\diezala#1{10$^{#1}$}

\def\Msun{\rm M_\odot}
\def\Eg{E_{\rm grav}}
\def\Mcl{M_{\rm cl}}
\def\Wext{W_{\rm ext}}

\def\be{\begin{equation}}
\def\ee{\end{equation}}

\def\mnras{MNRAS}
\def\apjs{ApJS}
\def\apj{ApJ}
\def\apjl{ApJL}
\def\araa{ARAA}

\shorttitle{Dark matter in cloud complexes}
\shortauthors{Su\'arez-Madrigal et al.}

\begin{document}

\title{Dark Matter as an active gravitational agent in cloud complexes} 
\author{Andr\'es Su\'arez-Madrigal\altaffilmark{1}, Javier Ballesteros-Paredes, Pedro Col\'in, and Paola D'Alessio}
\affil{Centro de Radioastronom\'ia y Astrof\'isica,
       Universidad Nacional Aut\'onoma de M\'exico, \\
            Apdo. Postal 72-3 (Xangari), Morelia,
            Michoc\'an, M\'exico. C.P. 58089 \\}

\altaffiltext{1}{a.suarez@crya.unam.mx}

\begin{abstract}
We study the effect that the dark matter background (DMB) has on the
gravitational energy content and, in general, on the star formation
efficiency of a molecular cloud (MC).  We first analyze the effect
that a dark matter halo, described by the \cite{NFW:96} density profile, has on
the energy budget of a spherical, homogeneous, cloud located at
different distances from the halo center.  We found that MCs located
in the innermost regions of a massive galaxy can feel a 
contraction force greater than their self-gravity
due to the incorporation of the potential of the galaxy's dark matter halo.
We also calculated analytically the gravitational perturbation that a
MC produces over a uniform DMB (uniform at the scales of a MC) and 
how this perturbation will affect the evolution of the MC
itself.  The study shows that the star formation in a MC will be
considerably enhanced if the cloud is located in a dense and low
velocity dark matter environment.  We confirm our results by measuring
the star formation efficiency in numerical simulations of the
formation and evolution of MCs within different DMBs.  Our study
indicates that there are situations where the dark matter's
gravitational contribution to the evolution of the molecular clouds
should not be neglected.
\end{abstract}


\section{INTRODUCTION}
\label{sec:intro}

A great effort from the astronomical community is currently directed
towards the detailed understanding of star formation, in particular to
the precise mechanisms involved in the conversion of some part of a
molecular cloud's mass into stars.
The evolution of molecular clouds (MCs) and the formation of stars
within them are thought to be regulated by actors as diverse as
gravity, large-scale flows, turbulence, magnetic fields, and stellar
feedback, among others \cite[see reviews by][]{SAL87,VS10}.

While evaluating the dynamical state of a molecular cloud due to all
relevant forces acting on it, the Virial Theorem is frequently invoked
to define an equilibrium condition by equaling the gravitational
energy $W$ of the cloud to twice its kinetic energy $K$:

\be
2 K = - W.
\ee

Traditionally, this relationship has been interpreted in terms of
clouds being in a state of quasi-equilibrium and long-lived entities,
and deviations from this condition are assumed to mean that the cloud
is collapsing or expanding, depending on which term is dominant
\citep[e.g.,][]{Myers_Goodman88a, Myers_Goodman88b, McKee_Zweibel92}.
However, it has been demonstrated that a cloud fullfiling the so-called
equilibrium condition will not necessarily remain dynamically stable 
\citep{BP06}. In fact,
numerical simulations show that clouds collapsing in a chaotic and
hierarchical way will develop such a relationship \citep[e.g.,
][]{VS+07, BP+11a, BP+11b}.

Furthermore, when calculating the Virial Theorem for MCs, it is a
common practice to consider that the gravitational term is the
gravitational energy, assuming that the cloud is, in practice, an
isolated entity. However, different recent works have showed that the
media surrounding the dense structures can importantly alter the
evolution of such entities \citep[e.g., ][]{BH04, Gomez+07, VS+07,
  HH08, BP+09a, BP+09b}.  In these works, the influence of barionic
matter outside the region of interest is studied.

On an apparently completely different topic, evidence from a wide
range of observations has long led astronomers to argue in favor of
the existence of much more matter than is actually seen: the elusive
dark matter.  Although the very nature of this mass component has not
been discovered yet, it is presumed to smoothly permeate the majority
of galaxies and in many cases to greatly surpass their visible
extension, while being their dominant mass component. 
Compared to the sizes of giant Galactic molecular clouds, dark matter
halos are huge and, for most practical purposes, homogeneous.
Thus, one would expect that such a distribution will not cause a
substantial gravitational effect on the cloud.

In the present contribution, the idea that dark matter halos may have
an effect on the energy content of the gas in galaxies is investigated.  
As a first approach, we focus our study in the center of massive galaxies,
where the effect is expected to be the greatest.  A cusped dark
matter profile such as the one proposed by \citet[][hereafter
  NFW]{NFW:96} would provide the central region of a halo with a
potential well which could importantly contribute to the gravitational
energy of a molecular cloud located around there.

In addition, we explore the idea that a barionic mass concentration is
able to introduce a perturbation in an otherwise homogeneous dark
matter distribution. As such, the mere presence of a molecular cloud
would modify the dark matter halo in which it is embedded.  This
perturbation would in turn place the cloud in a local external
potential well, which would play as an extra agent in the struggle
that determines its dynamical state. In this case, the external medium
to the MC will contribute to its collapse. When compared to a
situation identical to this one but without the dark mater background
(DMB), one would expect that a MC in this scenario would display an
enhanced capacity to form stars, since there is an extra factor
working in the same direction as the cloud's internal gravitational
energy. As such, measuring the star formation efficiency (SFE) of
clouds embedded in different DMBs would hint towards the importance of
the environment in their evolution in each case. We analytically study
the situation and give light as to what physical conditions would be
required for this to be an important effect. 

 Furthermore, we present
results from numerical simulations of molecular cloud formation and
evolution from two convergent monoatomic flows, placed in different
dark matter contexts.  Although the DMBs explored are not very
realistic, they are designed to test the analytical predictions and
illustrate the importance of accounting for the external gravitational
potential in molecular cloud analysis.

The structure of the article is as follows: in section
\S\ref{sec:Wext} we introduce the tidal energy term that is part of
the total gravitational energy of a matter distribution.  In section
\S\ref{sec:wexthalo} we describe a semi-analytical procedure to
evaluate the tidal contribution that the complete halo of a galaxy can
have on a molecular cloud and apply it to a few scenarios.  In section
\S\ref{sec:perturbationanalysis} we analytically derive an expression
to asess the tidal effect of the dark matter background perturbation
caused by the molecular cloud itself, while section
\S\ref{sec:semianalitical} evaluates this effect in different dark
matter background environments. In section \S\ref{sec:sims} we present
numerical simulations of cloud formation, in order to test the
analytical predictions in a more realistic scenario, and in
\S\ref{sec:numericalresults} we show the results obtained.  Finally,
\S\ref{sec:concs} gives general conclusions for the two studied
effects.

\section{FULL GRAVITATIONAL CONTENT OF A MOLECULAR CLOUD IN A GALAXY}
\label{sec:Wext}

Molecular clouds, like any other physical system, are bound to follow
all the internal and external forces acting on them and determining
their behavior over time.  Along the present article, the total
gravitational content for MCs situated in different environments will
be evaluated, in order to look for the effect that their surroundings
can have on their evolution.  Hence, this section derives an
expression for the tidal energy component, which will allow for an
evaluation of its gravitational effect on the cloud.

With the intention of determining a MC's dynamical condition, an
energy balance is commonly performed to it by invoking the Virial
Theorem.
In this theorem, the gravitational energy term $W$ considered for a
mass distribution of volume $V$ and density $\rho$, embedded in a
gravitational potential $\Phi$ is given by

\be
    W = -\int_V \rho\ x_i\ \frac{\partial \Phi}{\partial x_i}\ d^3x
\ee
\citep[see, e.g., ][]{SAL87}.  When applied to MCs, this term is
traditionally assumed to be equal to the gravitational energy of the
cloud:

\be
   E_{\rm grav} = \frac{1}{2} \int_V{\rho_{\rm cl}\ \Phi_{\rm
       cl} \ d^3x},
\label{eq:Eg}
\ee
where $\rho_{\rm cl}$ represents the density of the cloud and 
$\Phi_{\rm cl}$ the potential energy that it generates.
This assumption considers that the relevant gravitational potential is
due only to the mass of the cloud, which is a valid approximation for
an isolated mass distribution \citep{BP06}.  However, in a given
galaxy, there is a considerable amount of material (both barionic and
dark matter) coexisting with a MC. Thus, following \citet{BP+09a},
one may write the
gravitational potential $\Phi$ of a MC in a galaxy as the potential
due to the mass of the cloud itself, $\Phi_{\rm cl}$, plus the
contribution of any other external sources, $ \Phi_{\rm ext}$,

\begin{equation}
  \Phi = \Phi_{\rm cl } + \Phi_{\rm ext},
\label{eq:phi}
\end{equation}
such that the gravitational term entering the Virial theorem becomes

\begin{equation}
  W = \Eg + W_{\rm ext} .
\label{eq:Wtot}
\end{equation}
where 

\be
   W_{\rm ext} \equiv -\int_V \rho_{\rm cl}\ x_i\ \pd{\phi_{\rm ext}}{x_i}\ 
       dV,
	\label{eq:Wext}
\ee
is the tidal energy, i.e., the gravitational energy contained in the cloud,
but due to any mass distribution external to the cloud. As explained with
detail in \citet{BP+09a}, the external gravitational energy 
can either contribute to the collapse
of the cloud or to its disruption, depending on the concavity of the
external gravitational potential where it is located.

In the following sections, the tidal energy felt by a given MC will be
compared to its internal gravitational energy; its relative importance
will determine the role that any material surrounding a MC will play
in its evolution and future star formation.

\section{TIDAL ENERGY IN THE POTENTIAL WELL OF A DARK MATTER HALO}
\label{sec:wexthalo}
\begin{figure}
        \includegraphics[scale=0.3,angle=-90]{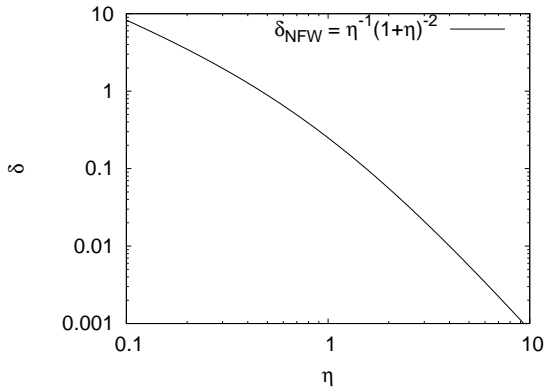}
        \caption{Dimensionless NFW density profile ($\delta_{\rm
            NFW}$) as a function of radius ($\eta$). The density has a
          cusp in the center of the matter distribution and diverges
          at $\eta = 0$. This profile appropriately describes dark
          matter halos found in numerical simulations.}
        \label{fig:densNFW}
\end{figure}

To evaluate the effect that dark matter might have on star
formation, we first consider a toy model consistent of a spherical
molecular cloud embedded in a spherical dark matter halo, with the
cloud located at a distance $s$ from the halo's center.  The halo
density profile is given by the simulation-inspired NFW density
profile \citep{NFW:96}:

 \be \rho_{\rm NFW}(r) = \frac{\rho_s}{(r/r_s)(1+r/r_s)^2},
        \label{eq:rhoNFW}
\ee
where $r_s$ is a scale radius, the radius where the logarithmic
derivative of the density is equal to $-2$, and $\rho_s$ is a
characteristic density, the density at $r = r_s$.  The profile can be
expressed in a dimensionless form by

\be \delta_{\rm NFW}(\eta) =
\eta^{-1}(1+\eta)^{-2},
        \label{eq:rhoNFWadim}
 \ee 
where $\delta_{\rm NFW}(\eta) \equiv \rho_{\rm NFW}(r)/\rho_s$ and
$\eta \equiv r/r_s$.  As seen in Fig. \ref{fig:densNFW} or eqs.
(\ref{eq:rhoNFW}) and (\ref{eq:rhoNFWadim}), the NFW profile has a cusp at the center.  It goes to infinity
as $\eta^{-1}$ for $\eta \rightarrow 0$.

The NFW gravitational potential can be expressed as

\be
        \phi_{\rm NFW}(r) = -4\pi G \rho_s r_s^2 \frac{\ln{(1+r/r_s)}}{r},
\ee
where $G$ is the gravitational constant. A dimensionless version of it
would be

\be
        \psi_{\rm NFW}(\eta) = \frac{\ln{(1+\eta)}}{\eta},
\ee
where $\psi_{\rm NFW} \equiv -\,\phi_{\rm NFW}/4 \pi G \rho_s r_s^2$
and $\eta \equiv r/r_s$.  It is noteworthy that, although the density
profile is divergent, the potential is finite everywhere due to the
mild divergence of the density; the limit of the NFW potential as the
radius goes to zero is 

\be 
  \lim_{r \to 0}{\phi_{\rm NFW}} = -4\pi G \rho_s r_s^2.  
\ee

We assume, for simplicity, a molecular cloud modeled as a sphere with
constant density $\rho_{\rm cl}$ and radius $R$.  When calculating
$W_{\rm ext}$ (eq. \ref{eq:Wext}), it is important to note that the
integration should be done over the volume of the molecular cloud,
while the expression for the NFW potential is centered on the dark
matter halo.  A coordinate transformation is then necessary to match
both systems of reference (separated by the distance $s$), which makes
the expression significantly more elaborate for any configuration
where the cloud's center doesn't coincide with the center of the halo
(i.e. $s \ne 0$).  For this reason, a numerical integrator needs to be
used to obtain the tidal energy.  

To get a quantitative idea of the importance of the contribution of
the dark matter halo, $W_{\rm ext}$ is compared to the internal
gravitational energy content of the cloud.  For a spherical and
homogeneous distribution, this amounts to

\be 
E_g = -\frac{16}{15}\pi^2 G \rho_{\rm cl}^2 R^5.
\label{eq:Egspherical}
\ee
The energy ratio $W_{\rm ext}/E_g$ measures the relative importance of
the gravitational energy contribution of the dark matter
compared to the gravitational energy of the cloud itself.

We have, in total, five parameters available to explore:
two NFW halo parameters (characteristic density and radius of the
density profile), two molecular cloud properties (density and extent
of the cloud), and the distance of the cloud from the center of the
halo ($s$).  We choose to fix first the halo characteristics and the
distance $s$ and calculate $W_{\rm ext}/E_g$ values for a range of
cloud's parameters.  We present energy ratio maps for typical
molecular cloud radii and densities, one per halo type and cloud
separation.

The gravitational effect of a Milky Way-type dark matter halo on a MC
is first presented.  As NFW parameters, we choose values predicted by
the $\Lambda$CDM cosmology with $\Omega_m = 0.3$ and $\Omega_\Lambda =
0.7$, for a halo of $M_{\rm vir} = 10^{12}~ M_\odot$ \citep[the
  estimated mass for our Galaxy;][]{B+05}.  For this mass, \cite{B+01}
predict a concentration $c = 7.9$, where $c \equiv R_{\rm vir} / r_s$.
The virial radius $R_{\rm vir}$ is defined as the radius where the
mean density of the halo is $\delta$ times the average density of the
universe. For the cosmology considered here $\delta = 337$ at $z =
0$. Moreover, $M_{\rm vir}$ is simply the mass inside $R_{\rm vir}$.  The
model by \cite{B+01} gives the median concentration for a given mass,
redshift and cosmology, based in statistical results from cosmological
simulations.  Once we have $M_{\rm vir}$ and $c$, the NFW profile is
completely specified; for example, using the definition of $R_{\rm
  vir}$ and the value of $c$ one can obtain $r_s$, while $\rho_s$ can
simply be computed by integrating the NFW profile from 0 to $R_{\rm
  vir}$ and using the values of $r_s$ and $M_{\rm vir}$.  Thus, for
the dark matter halo expected to permeate a Milky Way-type galaxy,
$\rho_s = 3.6\times10^{-3}$\,$\Msun$ pc$^{-3}$ and $r_s = 25.8$\,kpc.

\begin{figure}
        \includegraphics[scale=0.40]{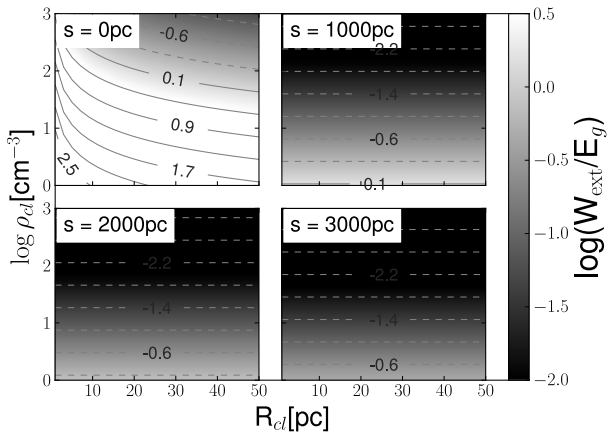}
        \caption{$W_{\rm ext}/E_g$ map for constant density clouds of
          radius $R_{\rm cl}$ and density $\rho_{\rm cl}$, located
          inside a Galactic-type dark matter halo ($M = 10^{12}\Msun$)
          and centered at a distance $s$ from the halo center. When
          the cloud center is located at the origin of the halo, its
          tidal energy can have a comparable effect to that of the
          gravitational energy. In general, the smaller and less dense
          clouds feel a greater gravitational contribution from the
          external medium.}
        \label{fig:especNFWVL}
\end{figure}

Figure \ref{fig:especNFWVL} shows the ratio $\log (W_{\rm ext}$/$E_g$)
in gray scale, as a function of the cloud's density ($y$-axis) and
radius ($x$-axis) for different values of the separation $s$ between the
cloud and halo's center.  On the top left panel, we show the case
where the test cloud is centered at the halo's center ($s=0$~pc).
For this particular setup, the tidal energy is quite important
relative to the gravitational energy of the cloud  (the energy
ratio is bigger than 0.2) for a wide range of cloud's densities and
radii: $\rho_{cl} \simless 10^3 \cmtres$ and $R_{cl} \simless$ 50 pc.
The top right panel shows a cloud placed 1~kpc away from the center of
the halo. The importance of the tidal energy has already decreased for
high density clouds of the same size.  As the cloud moves away from
the center of the halo ($s=2, 3$~kpc, lower panels), the tidal
contribution rapidly becomes smaller, as can be seen in the bottom
panels.  Here, the tidal contribution is negligible except for very
low density clouds.

Only clouds located very near the center of the halo feel a noticeable
influence from it: the values for the energy ratio decay rapidly as
the separation between the cloud and the halo increases.  This is due
to the central cuspiness of the NFW profile, shown in figure
\ref{fig:densNFW}, since $W_{\rm ext}$ is dependent on the potential
gradient which grows steeper near the origin.  It is also evident that
values of the tidal energy are more important for smaller and less
dense clouds.  Since the external gravity is ``pulling'' the particles
in the cloud, it can be understood that a cloud of smaller diameter
and lower density will be less tightly bound and easier to influence.
 
\begin{figure}
        \includegraphics[scale=0.40]{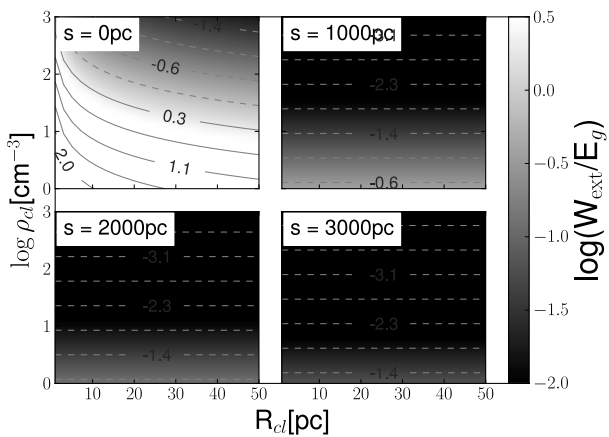}
        \caption{As in Fig. \ref{fig:especNFWVL} but for a galaxy with
          a halo mass of $10^{10}$\,$\Msun$.  A similar effect is
          noticed, but scaled down due to the potential well of such a
          halo being shallower.}
        \label{fig:especNFW1010}
\end{figure}

As a second example, we present the results from an identical estimate
but for a dark matter halo of mass $10^{10}$\,$\Msun$.
In this case, the NFW parameters employed are $\rho_s = 8.5
\times10^{-3}$\,$\Msun$ pc$^{-3}$ and $r_s = 3.8$\,kpc.  As in Fig.
\ref{fig:especNFWVL}, in Fig. \ref{fig:especNFW1010} we present the
$\log{( W_{\rm ext}/E_g)}$ maps for this halo's characteristics.  
We notice that 
for a halo with these parameter values the effect is even weaker, and
fades away faster as the molecular cloud moves away from the center of
the halo.  This smaller contribution can be explained as a result of
the shallower potential well of this less massive halo.  

It is important to mention that the ratios presented here only show the
importance of the dark matter tidal energy compared to the gravitational
energy of a cloud. We do not make any comparison with the gravitational
influence of other possible matter components, such as a galaxy's central
black hole or star population.
With the results shown, we conclude that a dark matter halo is only capable
 of influencing the evolution of small, diffuse clouds which are located 
near the center of massive galaxies.

\section{GRAVITATIONAL {\it FEEDBACK} FROM A PERTURBED HALO}
\label{sec:feedback}

\subsection{Perturbation Analysis}
\label{sec:perturbationanalysis}

To continue with another perspective of the dark matter influence in a
molecular cloud's evolution, this and subsequent sections study the 
gravitational {\it feedback} effect that a perturbation on the local
 dark matter background (DMB), produced by a barionic cloud, has upon 
the energy budget of the cloud itself.

Dark matter interacts with barionic matter only gravitationally; as
such, dark matter particles behave as a non-colissional fluid that
feels a gravitational attraction to a mass distribution like a
molecular cloud.  Because galaxies are immersed in dark halos, a MC
residing in it will coexist with a DMB.  Following a dark halo
distribution like that of NFW, this DMB would have an almost constant 
density at scales of the MC, because the variation scale of the halo's density
profile is much larger than the extent of even a giant molecular cloud
(GMC).  However, the mere presence of the barionic cloud perturbs the
otherwise homogeneous DMB and promotes a density enhancement in it,
which generates a potential well that will influence the MC in return.
We analyze how important this perturbation can be to alter the
evolution of a molecular cloud in different dark matter environments.

Although molecular clouds are highly irregular and structured
\citep[][and references therein]{Combes91, BP+07}, we model them as
Plummer spheres in order to follow up an analytical expression which
may be easy to use.  The mass distribution of such a sphere follows a
density profile given by

\be
  \rho_{\rm cl} = \frac{3\Mcl}{4\pi a^3} \del{1 +
  \frac{r^2}{a^2}}^{-5/2},
\label{eq:rho_P}
\ee
where $\Mcl$ is the total mass of the cloud, $r$ is the radial
distance from the center of the cloud, and $a$ is the scale-radius of
the Plummer potential, which gives an idea of the size of the core of
the distribution.  
The associated gravitational potential can be written as
\be 
   \Phi_{\rm cl}(r) = -\frac{G \Mcl}{a} \frac{1}{\sqrt{1+r^2/a^2}},
\label{eq:phi_P}
\ee
where G is the gravitational constant.
A cloud with these profiles is then placed within a DMB with constant
density $\rho_{\rm DM}$ and a velocity dispersion $\sigma_{\rm DM}$.

Following \cite{Hernandez_Lee08}, the dark matter particles are
described by a Maxwell-Boltzmann velocity distribution function
$f_0(v) \propto \exp(-v^2/2\sigma_{\rm DM}^2)$. For the purpose of
this study, the dark matter particles will have an average velocity of
0, that is, the DMB as a bulk is considered at rest with respect to the 
barionic matter.  Although the gravitational potential of the dark matter 
halo $\Phi_0$ is constant at the scales of the molecular cloud, the MC
induces a perturbation $\Phi_1$, so that the total potential $\Phi$ is
the addition of both contributions, i.e.  

\be 
  \Phi (r) = \Phi_0(r) + \epsilon \Phi_1(r), 
\label{eq:epot}
\ee
and the original density $\rho_0$ is enhanced by a perturbation
$\rho_1$ and should become 

\be \rho(r) = \rho_{0} + \epsilon \rho_1(r) .
\ee

Similarily, the original Maxwell-Boltzmann distribution function
$f_0(v)$ of the local dark matter halo gets perturbed by an amount
$f_1(r,v)$ and the new distribution function will be given by 

\be
f(r,v) = f_0(v) + \epsilon f_1(r,v).
\label{eq:ef}
\ee

In order to know how the dark matter potential will be affected by the
presence of the molecular cloud, we must solve the Boltzmann equation
for non-collisional systems
\be
    \pd{f}{t} + \textbf{v} \cdot \nabla f - \nabla \Phi \cdot \nabla_v
        f = 0,
\label{eq:NCBE}
\ee
in which $\nabla$ represents the gradient operator with respect to
spatial coordinates and $\nabla_v$ the gradient operator with respect
to velocity coordinates.  The quantities involved in eq.
(\ref{eq:NCBE}) are the ones taking into account the perturbation
caused by the barionic matter, i.e., the enhanced potential
(eq. \ref{eq:epot}) and distribution function (eq. \ref{eq:ef}).
For simplicity, we look for a first order stationary solution to the
Boltzmann equation (i.e. $\partial{f}/\partial{t} = 0$).  By using the
Jean's swindle ($\Phi_0 = 0$), the fact that $f_0$ does not explicitly
depend on the spatial coordinates ($\partial{f_0}/\partial{x_i} = 0$),
and considering only first order terms, the radial solution can be
obtained as

\begin{eqnarray}
   \textbf{v} \cdot \nabla f &=& \nabla \Phi \cdot \nabla_v f,
   \nonumber\\
   &=& \nabla (\Phi_0 + \Phi_1) \cdot \nabla_v(f_0 + f_1),\nonumber\\
   &=& \nabla \Phi_1 \cdot \nabla_v f_0 + \nabla \Phi_1 \cdot \nabla_v
   f_1,\nonumber \\
    &\approx& \nabla \Phi_1 \cdot \nabla_v f_0.
   \label{eq:radial}
\end{eqnarray}

We can calculate the gradient of the gravitational potential
perturbation $\Phi_1$, which is actually the cloud potential
$\Phi_{\rm cl}$, as 

\be
   \pd{\Phi_1}{r} = \frac{G \Mcl}{a^{3}} \frac{r}{(1+r^2/a^2)^{3/2}}.
\ee 
Using the explicit form of the Maxwell-Boltzmann velocity distribution
function to get the velocity gradient,

\be 
   \pd{f_0}{v} = -\frac{v}{\sigma_{\rm DM}^2} f_0(v), 
\ee
and after inserting the last two expressions in eq. (\ref{eq:radial}),
it becomes

\be 
  v \pd{f_1}{r} = \frac{rG\Mcl}{a^3 (1+{r^2}/{a^2} )^{3/2}}
     \del{-\frac{v}{\sigma_{\rm DM}^2}} f_0(v),
\ee
which furthermore, can be integrated over velocity space as 

\be
\pd{\rho_1}{r} = -G\Mcl \frac{r}{a^3(1 + {r^2}/{a^2} )^{3/2}}
\del{\frac{1}{\sigma_{\rm DM}^2}} \rho_{0}(v).  
\ee 
Remembering that the original DMB density has a constant value of
$\rho_{\rm DM}$, this equation can be solved to yield 

\be 
\rho_1 = \frac{G\Mcl}{a\ \sigma_{\rm DM}^2 } 
         \frac{1}{\sqrt{1+{r^2}/{a^2}}} \rho_{\rm DM},
\label{eq:rho_perturbed}
\ee
which is the density perturbation in the dark matter distribution
produced by the molecular cloud.  Such a density perturbation in the
dark matter halo will produce its own non-constant gravitational
field, which may, in principle, influence the gravitational energy budget 
of the molecular cloud.  Our interest lies, thus, in the gravitational
feedback of this enhancement over the cloud itself. 

In order to calculate the tidal energy produced by the perturbed dark
matter halo over the cloud, it is necessary to calculate the gradient
of the gravitational potential $\phi_{\rm DM}$ associated to the
perturbation $\rho_1$ given by eq. (\ref{eq:rho_perturbed}). The
gravitational potential as a function of the radius $r$ that is
produced by the perturbation $\rho_1$ will be given by

\begin{eqnarray}
   \phi_{\rm DM}(r) & =& -4\pi G \biggl\{ \frac{1}{r} \int_0^r
     \rho_1(r')r'^2 dr' \nonumber \\
    & & + \int_r^{R_{\rm DM}} \rho_1(r')r' dr' \biggr\},
\end{eqnarray}
integrated over the radial coordinate $r'$ and where the second
integral will be evaluated up to some radius $R_{\rm DM} $, such that
the perturbation is fully contained within this radius, i.e. $R_{\rm
  DM}$ is taken such that $\rho_1(r=R_{\rm DM}) = \rho_0$. Introducing
(\ref{eq:rho_perturbed}) and integrating, we get an expression for the
perturbed dark matter potential:

\begin{eqnarray}
    \phi_{\rm DM}(r)&=&-\frac{2\pi G^2 \Mcl a}{\sigma_{\rm DM}^2}\rho_{\rm DM}
    \biggl\{(r/a)^{-1}\ln{\del{r/a+\sqrt{1+(r/a)^2}}}\nonumber \\
    &&- \sqrt{1+(r/a)^2} + 2\sqrt{1+(R_{\rm DM}/a)^2}\biggr\},
\end{eqnarray}
whose gradient is given by 
\begin{eqnarray}
   {\partial{\Phi_{\rm DM}} \over \partial r} & = & -\frac{4\pi G^2 \Mcl
     a}{\sigma_{\rm DM}^2} \rho_{\rm DM} \biggl\{\frac{a \ln{\del{r/a +
     \sqrt{(r/a)^2 + 1}}}}{2r^2}\nonumber \\
   & & - \frac{\sqrt{1+(r/a)^2}}{2r}\biggr\} .
\end{eqnarray}
Introducing this equation and the Plummer density distribution
(\ref{eq:rho_P}) into eq.  (\ref{eq:Wext}), we obtain

\begin{eqnarray}
   W_{\rm ext} = &-&\frac{2\pi G^2 a \Mcl^2 \rho_{\rm DM} }{\sigma_{\rm DM}^2}
     \biggl\{\frac{2R/a}{1+(R/a)^2} - \arctan{\del{R/a}}\nonumber\\
    &-& \frac{\ln{\del{R/a+ \sqrt{1+(R/a)^2}}}}{(1+(R/a)^2)^{3/2}}\biggr\},
\label{eq:Wext_last}
\end{eqnarray}
where $R$ is the truncation radius of the Plummer profile for the
molecular cloud.  This expression quantifies the tidal energy of the
perturbed dark matter distribution over the cloud; in order to
evaluate its importance in the process, it should be compared to the
gravitational energy of the cloud itself, given by equation
(\ref{eq:Eg}).  Using eqs. (\ref{eq:phi_P}) and (\ref{eq:rho_P}) to
evaluate this expression, the gravitational energy of the cloud
becomes:

\begin{eqnarray}
E_{\rm grav} = -\frac{3G\Mcl^2}{2a} \biggl\{& &\frac{R/a}{8 \del{1+(R/a)^2}}
  - \frac{R/a}{4 \del{1+(R/a)^2}^2} \nonumber \\ 
   &+& \frac{1}{8} \arctan{\del{R/a}}\biggr\}.
\end{eqnarray}
Thus, ratio of the gravitational to the tidal energy is
\be
 \frac{W_{\rm ext}}{E_g} =  \del{\frac{a}{10 \rm pc}}^2
    \del{\frac{\rho_{\rm DM}}{\rm \Msun pc^{-3}}} 
  \del{\frac{\sigma_{\rm DM}}{\rm km\,s^{-1}}}^{-2} f(R/a), 
\label{eq:energyratio}
\ee
where we have defined 

\be
f(r/a) =  -1.8025 \ \   \frac{f_{\rm
       ext}(R/a)}{f_g(R/a)}, 
\label{eq:fratio}
\ee
with

\begin{eqnarray}
f_{\rm ext}(R/a) = \biggl\{\frac{2R/a}{1+(R/a)^2} -
     \arctan{\del{R/a}}\nonumber\\
    - \frac{\ln{\del{R/a+ \sqrt{1+(R/a)^2}}}}{(1+(R/a)^2)^{3/2}}\biggr\} ,
\label{eq:fext}
\end{eqnarray}
and

\begin{eqnarray}
{f_g(R/a)} = \frac{R/a}{8 \del{1+(R/a)^2}}
  - \frac{R/a}{4 \del{1+(R/a)^2}^2} \nonumber\\
   + \frac{1}{8} \arctan{\del{R/a}}\biggr\}  .
\label{eq:fg}
\end{eqnarray}

Equation (\ref{eq:energyratio}) tells how important is a given dark
matter environment on the gravitational energy budget of a molecular cloud.
Several points must be stressed from this equation.  First of all, the
ratio of tidal to gravitational energy does not depend on the mass of
the cloud, neither on its density.  Furthermore, the tidal energy on
the cloud produced by the perturbed halo will be more important for
larger clouds, and larger background halo densities.  On the other
hand, for a halo's larger velocity dispersion $\sigma_{\rm DM}$, the
perturbation must be smaller and, as a consequence, the tidal energy will
also be smaller, when compared to the gravitational energy, as showed by the
$\sigma_{\rm DM}^{-2}$ dependence in eq. (\ref{eq:energyratio}).

\begin{figure}
  \includegraphics[scale=0.55, angle=0]{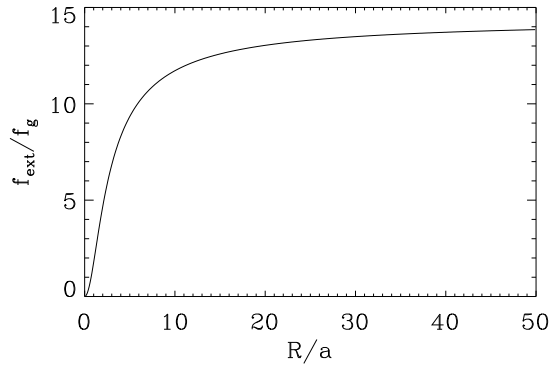}
  \caption{Ratio $f = f_{\rm ext}/f_g$ for a Plummer cloud of 
    truncation size $R$ and characteristic scale $a$ = 10~pc.  Note
    that the ratio is always positive, implying thus that the tidal
    energy always contributes to the collapse, according to
    eq. (\ref{eq:energyratio}.)
	\label{fig:fratio}}
\end{figure}

In Fig. \ref{fig:fratio} we plot the ratio $f$ vs $R/a$, as given by
eqs. (\ref{eq:fratio}--\ref{eq:fg}).  Since the function is always
positive, we note that the ratio of tidal to gravitational
energy, eq. (\ref{eq:energyratio}), will also be positive for any
physically consistent combination of parameters $R/a$, $\rho_{\rm
DM}$, $\sigma_{\rm DM}$.  This means that the tidal energy caused by
such a DMB will always contribute to the collapse of the MC perturbing
it.  This effect, when relevant, is expected to be reflected on the
evolution of the cloud. In a molecular region forming stars, a rise in
the compressional forces acting on it would show an increased star
formation efficiency (SFE), as more stars will form faster when there
is relatively less support against collapse.  In the following section
we explore the parameter space relevant to this scenario ($R/a$,
$\rho_{\rm DM}$, $\sigma_{\rm DM}$).

\subsection{Semi-analytical Results}
\label{sec:semianalitical}

In order to get an idea of the magnitude of the ratio $W_{\rm
  ext}/E_g$ in different dark matter environments, we need to provide
  dark matter backgrounds to the molecular clouds under study.  As in
  section \S3, we make use of the NFW model to estimate the structural
  parameters of the dark matter halos, where the MCs will be embedded.
  We need to determine the halo's density $\rho_{\rm DM}$ and velocity
  dispersion $\sigma_{\rm DM}$, given its total mass $M_{\rm vir}$ and
  the radius $r$ at which these properties are evaluated, at a
  specified redshift $z$. We only consider here relatively nearby
  galaxies; that is, we assume redshift $z = 0$.

To find $\rho_{\rm DM}$ and $\sigma_{\rm DM}$ we thus proceed as
follows: for a given halo mass $M_{\rm vir}$, we obtain the
concentration parameter $c$ following \cite{B+01}.
With these values, we calculate its density profile parameters as
specified in section \S\ref{sec:wexthalo}.  Once specified, we can
evaluate the profile at the desired radius $r$ to know the local
density $\rho_{\rm DM}$.  On the other hand, in order to obtain
$\sigma_{\rm DM}$, we adopt the relation given by \citet[][see their
  eq. 14]{LM01} for the case where velocities of dark matter particles
are considered isotropic.
Such expression is dependent only on $c_{\rm vir}$, $M_{\rm vir}$ and
$R_{\rm vir}$ and so the velocity dispersion obtained for a given halo
is a constant.  These recipes are used in our analysis to get the DMB
parameters needed in different cases.

To begin with, we first choose parameters consistent with a Milky-Way
type dark matter halo evaluated at a radius of 8 kpc, adequate to the
solar neighborhood.
However, in this case we preferred to
use a more realistic concentration parameter $c=12$, closer to the
observationally estimated value by, e.g., \citet{B+05}.
With these conditions, the procedure just described predicts a DMB with density
$\rho_{DM}\sim0.01 \Msun$\,pc$^{-3}$ and a velocity dispersion
$\sigma_{\rm DM} = 165$~km\,s$^{-1}$. 
With these parameters, equation (\ref{eq:energyratio}) predicts that
in order to get a non-negligible tidal effect ($W_{\rm ext}/E_g \sim
0.15$) the cloud's scale radius $a$ has to be huge ($a=1900$\,pc).  A
cloud of this size greatly surpasses the extent of local molecular
clouds.
  On the other hand, since the energy ratio $\Wext/\Eg$ varies
as $a^2$, we note that for more typical clouds, $a \sim 10$~pc and
$R/a\sim 10$, the contribution of the tidal energy is negligible.

\begin{figure}
  \includegraphics[scale=0.35, angle=-90]{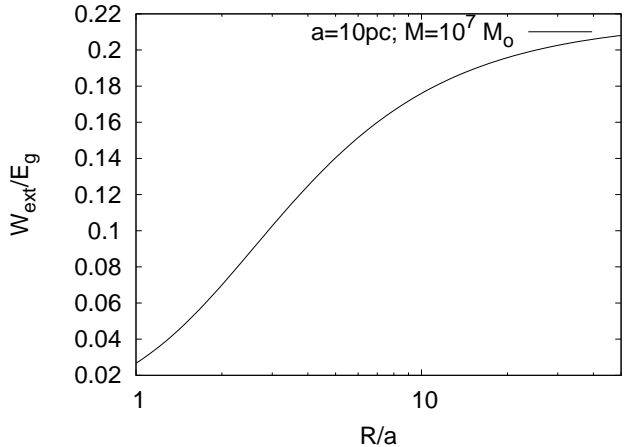}
  \caption{$W_{\rm ext}/E_g$ ratio for a cloud of Plummer radius $a =
    10$~pc and different truncation radii (R), embedded in a dark
    matter halo typical of dwarf galaxies ($M\sim 10^{7}\Msun$), at a
    radius $r=0.1r_s$. In this case, even small molecular clouds would
    feel a a non-zero tidal effect from their environment.}
	\label{fig:ratio_107}
\end{figure}

Figure \ref{fig:ratio_107} presents a more interesting case in which
the energy ratio is non-negligible for a range of cloud sizes
$R/a$. The cloud has a Plummer radius of $a=10$~pc and it is embedded
in a dark matter halo with mass $M=10^7\,\Msun$ at a distance of
$r=0.1r_s$.
We can see, in this case, that even small clouds of the order of 10\,pc 
will feel a non-zero tidal effect from the dark matter environment.

\begin{figure}
  \includegraphics[scale=0.45]{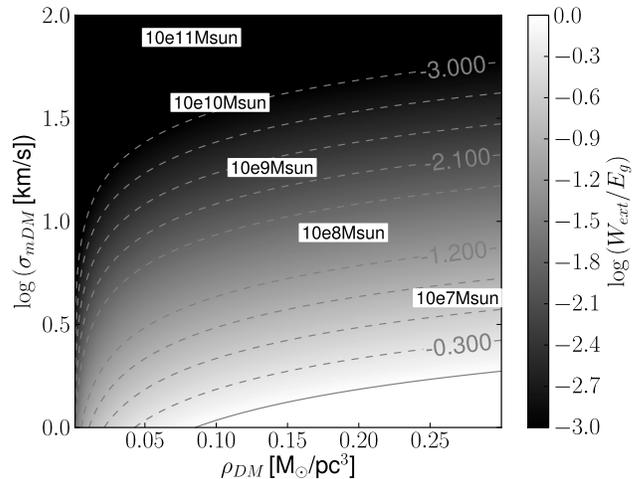}
  \caption{$W_{\rm ext}/E_g$ map for Plummer clouds with
    characteristic radius $a=10$~pc and $R/a$=10, embedded in a
    homogeneous dark matter background of parameters $\rho_{\rm DM}$
    and $\sigma_{\rm DM}$. The tidal energy can contribute to the
    collapse of clouds immersed in dwarf galaxies. As the energy ratio
    increases as $a^2$, also massive galaxies can show this effect for
    large enough MCs.}
	\label{fig:especRa10}
\end{figure}

\begin{figure}
  \includegraphics[scale=0.450]{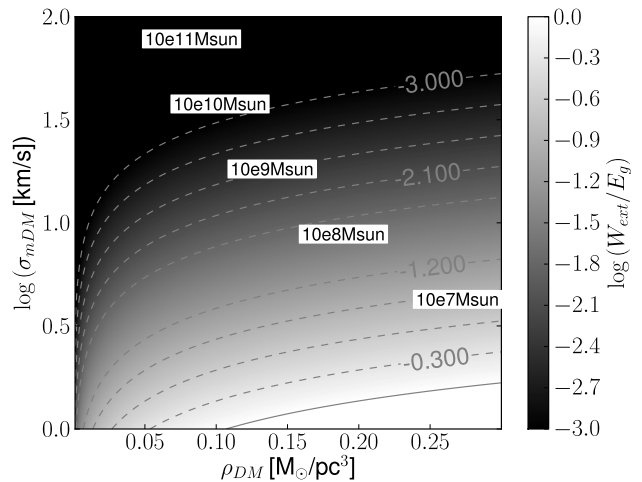}
  \caption{Same as in Fig.~\ref{fig:especRa10}, but for $R/a$=5. The
    differences with Fig.~\ref{fig:especRa10} are not significant
    because Plummer spheres contain most of their mass within 5
    characteristic radii.}
	\label{fig:especRa5}
\end{figure}

\begin{figure}
  \includegraphics[scale=0.450]{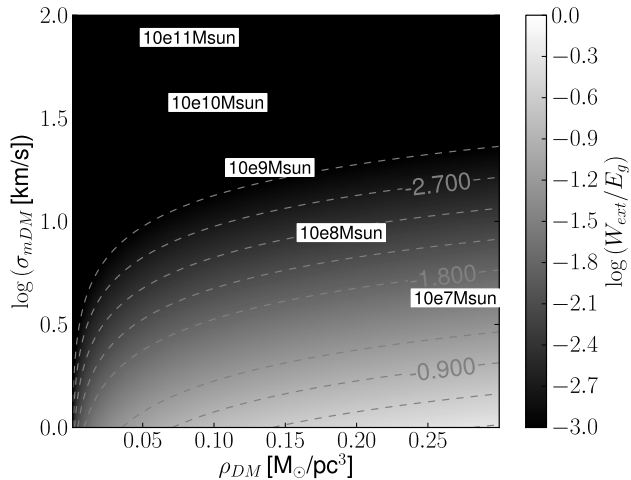}
  \caption{Same as in Fig.~\ref{fig:especRa10}, but for $R/a$=1. In
    this case, the tidal effect is diminished when compared to
    Fig.~\ref{fig:especRa10} because we are considering a more
    concentrated sphere in which the tidal forces have more
    difficulties to act.}
	\label{fig:especRa1}
\end{figure}

 Figure \ref{fig:especRa10} shows, in grey scale, the energy ratios
 $W_{\rm ext}/E_g$ evaluated using a NFW dark matter density profile
 with a range of velocity dispersions $\sigma_{\rm DM}$ and
densities $\rho_{\rm DM}$.  In this figure, the parameters of the 
Plummer cloud are $a=$10~pc, and $R/a=10$.  Similar maps are presented in
 Figs.~\ref{fig:especRa5} and \ref{fig:especRa1}, but for clouds with
 $R/a = 5$ and 1, respectively.  As a reference, in these images we
 also indicate the typical densities and velocity dispersions of dark
 matter halos of galaxies with total masses ranging from $10^7$ to
 $10^{11}~\Msun$. Density profiles are computed at $r=0.1r_s$.

Two points can be highlighted from Fig.~\ref{fig:especRa10}. First,
the tidal energy of the dark matter halo becomes important for large
clouds ($R\sim 100$~pc) in galaxies with masses ranging from \diezala
7 to \diezala 8~$\Msun$.  Secondly and as noticed previously, since the
ratio $\Wext/\Eg$ scales as $a^2$, this contribution become important
also in larger galaxies for large enough clouds.  For instance, as can
be inferred from Fig.~\ref{fig:especRa5}, $\sim$~10\% of the total
gravitational energy content of a giant cloud complex ($a\sim100$~pc,
$R\sim$~0.5~kpc) embedded in a \diezala{10}~$\Msun$ dark matter halo
will be due to the tidal compression energy produced by the potential
well induced by the cloud itself. 

 The high similarity between
Figs. \ref{fig:especRa10} and \ref{fig:especRa5} is likely due to the
mass of a Plummer sphere being concentrated in the inner 5
characteristic radii (93\% of the mass of the distribution lies at $r
< 4 a$).  On the other hand, Figure \ref{fig:especRa1}, which shows the
case $R/a = 1$, does show smaller contributions from the environment
to the energy budget of the cloud than those shown in previous
figures.  The reason can be traced again to the inherent concentration
of a Plummer sphere: as we truncate the sphere at $r = a$, we are only
considering the inner core of the cloud, which has a higher
gravitational energy and in which the tidal contribution has more
trouble to act.

\begin{figure}
  \includegraphics[scale=0.45]{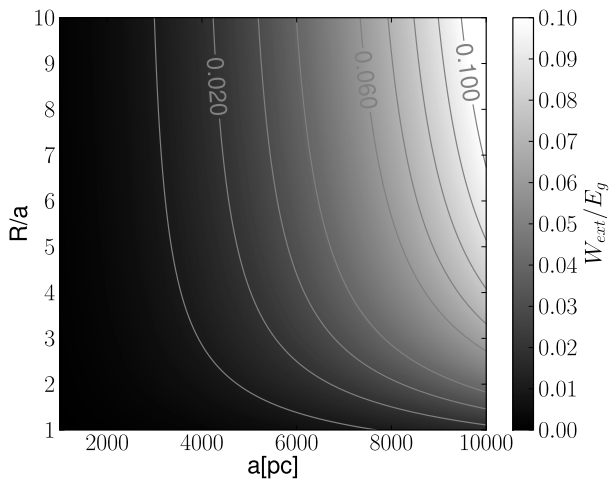}
  \caption{$W_{\rm ext}/E_g$ map for Plummer clouds of characteristic
    radius $a$ and truncation radius $R$, embedded in a homogeneous
    dark matter background with density and velocity dispersion
    corresponding to a Milky Way dark matter halo at 50~kpc from its
    center ($\rho_{\rm DM}=0.0002~\Msun$pc$^{-3}$, $\sigma_{\rm DM} =
    150$~km\,s$^{-1}$). A big mass distribution (like a satellite galaxy)
    would feel a small tidal energy contribution.}
        \label{fig:espec50kpc}
\end{figure}

To further study the ``big cloud'' scenario, a different exploration
 of parameters is shown in Figure \ref{fig:espec50kpc}, in which the
 density and velocity dispersion of the dark matter background are
 set according to the theoretical values of a Milky Way-type halo,
 at a radius of 50~kpc ($\rho_{\rm DM}=0.0002~\Msun$pc$^{-3}$, 
$\sigma_{\rm DM} = 150$~km\,s$^{-1}$), while the 
cloud characteristic radius $a$ and truncation radius $R$ are varied.
This allows us to identify the characteristics of a mass distribution
that would suffer a relevant contraction by the tidal forces: the
figure shows that, for a radius larger than $\sim8000$~pc
($a=8000$~pc, $R/a=1$, or $a=6000$~pc, $R/a=1.3$)
and located at some 50~kpc from the center of the Milky Way, it
will produce a perturbation in the dark matter halo which will in turn
have a small contribution to the gravitational energy it contains 
(around 1\%).  This location and dimensions are similar to that of a
satellite galaxy like the Large Magellanic Cloud.  The effect is
indeed low, but the theoretical implications it suggests are
interesting, as we could say that at least some contribution to the
star formation seen in orbiting galaxies or in galaxy mergers could
come from the tidal energy caused by a dark matter perturbation.\\

\begin{figure}
  \includegraphics[scale=0.45]{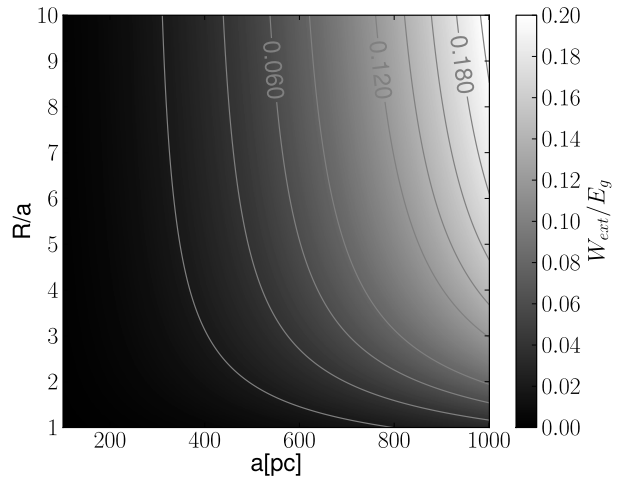}
  \caption{Same as in Fig. \ref{fig:espec50kpc}, but for a Milky Way
    dark matter halo at 2~kpc from its center ($\rho_{\rm
      DM}=0.04~\Msun$pc$^{-3}$, $\sigma_{\rm DM} = 150$~km\,s$^{-1}$). In this
    case, a large cloud complex would suffer a tidal effect from the
    dark matter, compared to the previous case.}
        \label{fig:especMW2kpc}
\end{figure}

Still, one more scenario to consider is that of a cloud closer to the
center of a galaxy Milky-Way type.  In Figure \ref{fig:especMW2kpc},
the dark matter parameters are fixed to fit the Milky Way's dark
matter halo at 2~kpc from its center ($\rho_{\rm
  DM}=0.04~\Msun$pc$^{-3}$, $\sigma_{\rm DM} = 150$~km\,s$^{-1}$), while the
cloud dimensions are explored.  Note that in this case, a large cloud
complex of $a\sim 1$~kpc and $R\sim 2$~pc will have a tidal energy
($\sim 6\%$) that will slightly contribute to its collapse.  This
possibility suggests that a complex like the molecular ring in the
Mily Way may be affected somehow by the tidal energy produced by the
perturbation in the halo that the cloud complex imprints on it.

\section{NUMERICAL SIMULATIONS OF CLOUD FORMATION WITHIN A DM HALO}
\label{sec:sims}

Different studies support the idea that molecular clouds in the Milky
Way are transient entities produced by compressions from large scale
streams in the interstellar medium \citep{BHV99, BVS99, HBB01, VS+07,
HH08, VS+10}. On the other hand, in sections
\ref{sec:perturbationanalysis} and \ref{sec:semianalitical} we derived
semi-analytical calculations which suggest that a dark matter well
produced by the perturbation of a barionic distribution of mass can
significantly add to its gravitational budget, and thus contribute to
its collapse, if the right conditions are given. In order to test this
scenario within a more realistic setting, we have performed a
series of numerical experiments in which a dense cloud is formed by
the collision of two streams of diffuse gas.
The colliding gas cools down, becomes denser and collapses. While
doing so, the collapsing cloud perturbs the locally homogeneous dark
matter halo\footnote{As the scale radii of the halos considered in
these simulations are much greater than the size of the clouds, this
is a good aproximation.}  in which it is embedded, producing a
potential well. This will give an extra tidal energy contribution to 
the original cloud, which is expected to translate into a larger star
formation activity.
Given the differences between the semi-analytical and the numerical
circumstances (initial conditions and treatment), we do not expect
semi-analytical results to faithfully reflect the behavior of the gas
and dark matter in the simulations;
still, the gravitational feedback effect must appear in the simulations. 
This effect can
be then estimated in the numerical experiments through the comparison
of the cloud's star formation efficiency with and without dark matter.

To perform the simulations, we use the Adaptive Refinement Tree
\citep{KKK97, Kravtsov03} code (ART), which employs an 
$N-$body algorithm to solve the gravitational interactions of the
system.  The chosen version of the code additionaly solves gas
hydrodynamics and can handle radiative cooling and star formation, as
presented by \citet{VS+10}.

The initial conditions for the gas are taken from previous star
formation simulations by \cite{VS+10}.  We consider a 256\,pc-per-side
box filled with tenuous gas ($n_0 = 1$\,cm\alamenos3) at a temperature
of $T_0 =5000$\,K, which represents a warm neutral hydrogen medium
(molecular weight $\mu_0 = 1.27$).  The gas has an inital turbulent
velocity fluctuation distribution of magnitude $v_{\rm rms} =
0.1$\,km\,s$^{-1}$\, everywhere.  On top of that, two gas streams moving
towards each other along the $x-$axis are created by adding an initial
velocity component of 7.5\,km\,s$^{-1}$, to the material contained in two
112\,pc long cylinders with radii of 64\,pc, lying on each side of the
box. Since the sound speed for the gas amounts to $c_s = 7.4$\,km\,s$^{-1}$,
the flows are trans-sonic. The gas that
comprises the streams constitutes less than 20\% of the total mass in
the box. The cloud evolution is followed during 40~Myr.

The local dark matter background is represented by particles that
interact only gravitationally, both with other dark matter particles
and with the gas. Table \ref{tab:runs} shows in the first and second
column the values of the density $\rho_{\rm DM}$ and the velocity
dispersion $\sigma_{\rm DM}$, respectively, used in each simulation.
These were calculated following the schemes described in section
\ref{sec:numericalresults}.  In the third column, we show the typical mass
of a halo (galaxy) associated with those $\rho_{\rm DM}$ and
$\sigma_{\rm DM}$ values, given at the radius shown in the fourth
column.


\begin{deluxetable}{ccccc}
\tablecolumns{5}
\tablewidth{0pt}
\tablecaption{Variable parameters in the simulations}
\tablehead{
    \colhead{$\rho_{\rm DM}$}				&
    \colhead{$\sigma_{\rm DM}$}				&
    \colhead{log($M_{\rm vir}$} 			& 
    \colhead{$r/r_s$} 					& 
    \colhead{Linestyle}      			       \\
    \colhead{($\Msun$pc$^{-3}$)} 			& 
    \colhead{(km\,s$^{-1}$)}           			&
    \colhead{$\rm [\Msun]$)}				&
    \colhead{}						&
    \colhead{in Fig. \ref{fig:simsSFE}}						}
\startdata
0 & 0 &(...)$^a$ & (...)$^a$&solid\\
0.1 & 37 & 10 & 0.15 & dashed \\
0.1 & 10 & 8 & 0.26 & long-dashed \\
0.1 & 2 & (...)$^b$ & (...)$^b$ & dash-dotted\\
0.1 & 1 & (...)$^b$ & (...)$^b$ & dotted\\
0.17 & 3.16 & 7 & 0.13 & dot-long dashed \\
0.24 & 3.16 & 7 & 0.1 & short-log dashed\\
\enddata
\tablecomments{Dark matter density
  ($\rho_{\rm DM}$) and vel. dispersion ($\sigma_{\rm DM}$) are shown,
  as well as the mass of a galaxy ($M_{\rm vir}$) whose characteristic
  NFW halo represents these values at the radius ($r$) given, when
  applicable.}
\tablenotetext{a}{Reference run}
\tablenotetext{b}{There are no realistic NFW dark matter
  halos with this characteristics}
\label{tab:runs}
\end{deluxetable}

To decouple the effect of the dark matter background on the
behavior of the gas from other effects, we need to study its evolution
without a dark matter background.  This reference run was
taken from \cite{VS+10} (SAF0, which stands for Small Amplitude
Fluctuations without stellar feedback) and it has the same values of the gas
parameters described above.  These authors report the formation of a
``central cloud'' of irregular shape with a radius of the order of
10\,pc, where star formation is concentrated. Although the physical
configuration of the simulation differs from the one of the
analytical case we studied previously, we believe it is a good test
case for investigating the general effect: the promotion of a higher
star formation rate.

In order to quantify the differencees in the evolution of the
different runs, we calculated the Star Formation Efficiency (SFE) as
defined by, for example, \citet{VS+10}:

\be 
{\rm SFE} = \frac{M_*}{M_*+M_{\rm gas}},
\label{eq:efe}
\ee 
where $M_*$ represents the mass contained in stars and $M_{\rm gas}$
accounts for the mass in the dense gas ($n \simgreat 10^2\ \cmtres$),
at any given moment during the evolution of the system.  We also
compute the relative difference between the SFE of any run with dark
matter and the reference run (without it), at any given time,
$\Delta$; that is,

\be
\Delta = \frac{{\rm SFE}_{\rm ref} - {\rm SFE}_{\rm run}}{\rm SFE_{\rm
    ref}}.
\label{eq:delta}
\ee

\begin{figure}
  \includegraphics[scale=0.6]{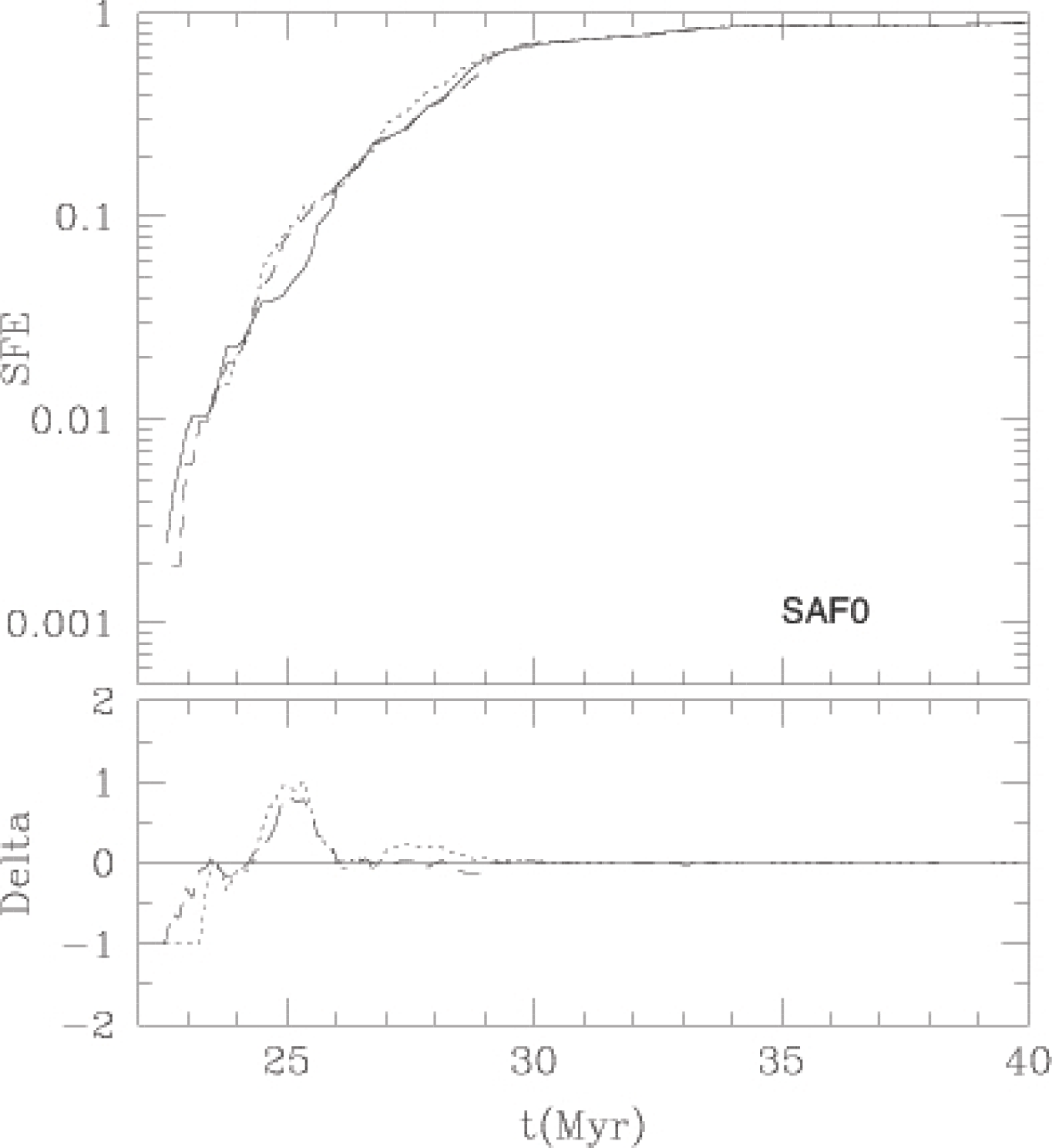}
  \caption{ SFE and differential values (with respect to the solid line run) 
    for three different runs with different random phases of the initial 
    density and velocity field,  but otherwise identical physical conditions:
    $\rho_{\rm DM} = 0.24\Msun$pc$^{-3}$,$\sigma_{\rm DM} = 3.16$\,km\,s$^{-1}$.
    As can be noticed, the differences are not large. Thus, for runs with 
    dark matter halo, any difference substantially larger than these shown
    in this figure will be considered as representative of the
    gravitational influence of the dark matter halo.}
    \label{fig:comparison}
\end{figure}

In this context, the first point to clarify is whether these
variables allow us to distinguish between the evolution of runs
with different physics (e.g., no dark matter included vs. a
particular configurations of dark matter halos), and the evolution of
runs with the same physics, but different initial fluctuations.
 Thus, in Fig.~\ref{fig:comparison}, we show the time evolution of the
 SFE (upper panel) for three runs with identical physics
 (with dark matter properties $\rho_{\rm DM} = 0.24\Msun$pc$^{-3}$, 
$\sigma_{\rm DM} = 3.16$\,km\,s$^{-1}$), but with different phases 
for the initial density and velocity fluctuations. 
However, in this case and unlike explained before, what is drawn 
as $\Delta$ in the lower panel of Fig.~\ref{fig:comparison} is the relative 
difference between the SFE of 
the run plotted with a solid line\footnote{The seed used to run this model is 
the same used to run the models with dark matter shown in Table \ref{tab:runs}} and the runs plotted with a dashed or a dotted line (which differ only in
the phase of the initial conditions).
 As can be seen in this figure, the differences among the three runs are 
not large, so the different initial conditions do not constitute a 
significant source for variations in the SFE. 
 Thus, if any of our performed runs with dark matter produces   
 substantially larger differences, it must be due mainly to the
 gravitational influence of the dark matter.

\section{NUMERICAL RESULTS} 
\label{sec:numericalresults}
\begin{figure*}
  \includegraphics[scale=0.8]{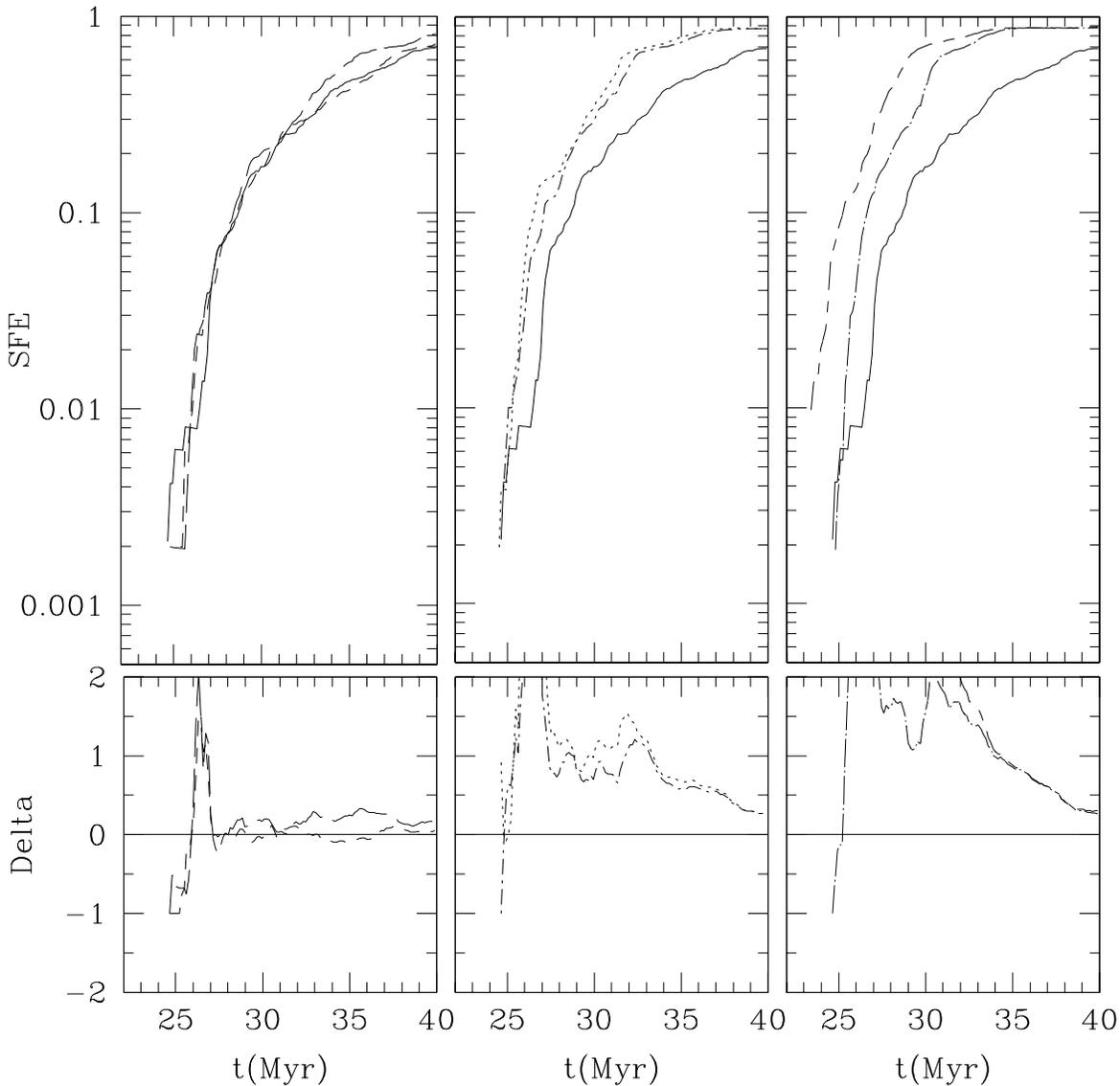}
  \caption{ SFE and $\Delta$ values for the simulation runs described
    in Table \ref{tab:runs}. The reference run is displayed as the
    solid line in all three panels.  The left panel compares high
    velocity dispersion DMBs ($\sigma_{\rm DM} \ge 10$\,km\,s$^{-1}$) with the
    reference run. Both runs in this panel have a density of
    $\rho_{\rm DM} = 0.1\Msun$pc$^{-3}$: the dashed line shows the 
    run with $\sigma_{\rm DM} = 37$\,km\,s$^{-1}$, while the
    long-dashed line depicts the one with $\sigma_{\rm DM} = 10$\,km\,s$^{-1}$.
    No significant difference is found from the reference run, as
    expected from the analytical predictions. The central panel shows
    two more runs with the same density ($\rho_{\rm DM} =
    0.1\Msun$pc$^{-3}$) but smaller velocity dispersions: $\sigma_{\rm
    DM} = 2$\,km\,s$^{-1}$ (dash-dotted line) and $\sigma_{\rm DM} = 1$\,km\,s$^{-1}$
    (dotted line).  These cases show a systematically higher SFE
    than the reference run, also as expected.  Two more runs with
    higher densities are displayed on the right panel: $\rho_{\rm DM}
    = 0.17\Msun$pc$^{-3}$, $\sigma_{\rm DM} = 3.16$\,km\,s$^{-1}$ (dot-long
    dashed line) and $\rho_{\rm DM} = 0.24\Msun$pc$^{-3}$,
    $\sigma_{\rm DM} = 3.16$\,km\,s$^{-1}$ (short-long dashed line). Again,
    the star formation seems a lot more active than the reference
    run.}
    \label{fig:simsSFE}
    \end{figure*}

As commented above, the collision of the streams in the simulations develops
instabilities in the atomic gas which changes its phase and becomes molecular
material with lower temperatures and higher densities; this new phase
forms condensations that
grow with time.  As the shocked region increases its mass, the cloud
starts contracting. The star formation starts at $\sim$~25\,Myr after
the begining of the simulation, and has a rapid growth soon
after. Around 10~Myr later, most of the gas in the simulation box has
been transformed into stars and hence the SFE approaches its maximum
possible value of 1 (see Fig. \ref{fig:simsSFE}).

In Figure \ref{fig:simsSFE} we show the time evolution of the SFE
(upper panels) and $\Delta$ (lower panels) for the different runs
outlined in Table \ref{tab:runs}, as compared to the reference run, which
is shown as a solid line in all panels.
On the left side we present the runs with 
density $\rho_{\rm DM} = 0.1\Msun$pc$^{-3}$ and velocity dispersions
$\sigma_{\rm DM} = 37$\,km\,s$^{-1}$ (dashed line) and $\sigma_{\rm DM} = 10$\,km\,s$^{-1}$
(long-dashed line).
Aside from the initial fluctuations, these runs show no significant 
difference in the evolution of the SFE, as compared to the reference 
run (solid line).
  This behavior is consistent with the results predicted for $\Wext/\Eg$
  by our semi-anlytical approach (see Figs. \ref{fig:especRa10},
  \ref{fig:especRa5} and \ref{fig:especRa1}). As far as the SFE is
  concerned, these simulations proceeded as if the dark matter was not
  present.
    The central panel depicts two more runs with the same density
    ($\rho_{\rm DM} = 0.1\Msun$pc$^{-3}$) but with lower velocity
    dispersions: $\sigma_{\rm DM} = 2$\,km\,s$^{-1}$ (dash-dotted line) and 
    $\sigma_{\rm DM} = 1$\,km\,s$^{-1}$ (dotted line).
    In this cases, there is a marked difference in the overall
    evolution of the SFE.
    The $\sigma_{\rm DM} = 2$\,km\,s$^{-1}$ run distinguishes from the previous 
    ones in that its SFE is
    consistenly higher than the one of the reference; there is a period of
    time in which it has an SFE a factor of two higher than the reference value.
    The dotted line in the
    same panel has an SFE that is systematicaly higher than all
    previous runs.  It can clearly be seen that runs with lower dark
    matter velocity dispersions (density remaining constant) produce higher
    SFEs.  As can be deduced from Figure \ref{fig:especRa5}, an important
    contribution is expected from the external gravitational potential
    towards the collapse of the cloud: a ratio $W_{\rm ext}/E_g$ in the
    range ~0.3-0.8 for these cases. This translates into a significant
    enhancement of the star formation rate.
    The right panel, in the same Fig. \ref{fig:simsSFE}, shows two other
    cases. These runs have both a higher density than those runs
    considered in the left and middle panels.  One run has $\rho =
    0.17\Msun$pc$^{-3}$ (dot-long dashed line) and the other one $\rho =
    0.24\Msun$pc$^{-3}$ (short-long dashed line). They are simulated with
    the same velocity dispersion $\sigma_{\rm DM} = 3.16$~km\,s$^{-1}$.  These
    models also show an enhanced SFE in comparison to the reference run,
    but here the contribution due to the DMB seems somewhat larger than
    the expected from the analytical predictions (see Figs.
		    \ref{fig:especRa10}, \ref{fig:especRa5} and \ref{fig:especRa1}).  Such
    a deviation is, however, not surprising: the predicted external
    contribution comes from an ideal model that considers a spherical
    molecular cloud at rest, while the simulation contains more realistic
    conditions.

    \section{DISCUSSION AND CONCLUSIONS}
    \label{sec:concs}

    We investigated the effect that the gravitational contribution from 
    dark matter, in the form of tidal energy, has on the energy balance of a
    molecular cloud and its evolution. On considering the large-scale
    dark matter halos, where galaxies are supposed to be immersed, we
    noticed that there is an important gravitational contribution only for
    clouds very near to the center of these halos, specially for low-density,
    small molecular clouds.  The effect of this collapse-promoting force
    dissapears very rapidly as the cloud location is moved away from the
    center. To our knowledge, this effect has not been studied before,
    perhaps in part because clouds were not expected to be affected by the
    presence of the dark matter.
    In light of these results, these predictions can be applied to
    structures encountered in the central region of galaxies.  In fact,
    \cite{G+:90} report on a series of observations regarding the rich gas
    structure in the inner 10~pc of the Milky Way, comprised of a couple
    giant molecular clouds, a circumnuclear disk and an HII region.  These
    entities lie at the right position, are small and clumpy (averaging
		    $\sim 3$~pc in diameter) and span a large range of densities
    \citep{VD:02}.  Such gas distributions can, according to our results,
    be feeling a great gravitational compression from the external
    medium, relative to their self-gravity.  
    These results also suggest
    that the intense star formation activity found in the innermost
    regions of large galaxies (like ours) may have some contribution from
    tidal compression caused by the dark matter halo of the galaxy.

    In this particular case,
    the presence of the central black hole \citep[with an estimated mass 
    $M_{\rm BH}=4\times10^6\Msun$;][]{S+02} is only relevant when the molecular
    cloud structure reaches distances smaller than $\sim 1$\,pc, in accordance 
    with the radius of influence of a central black hole $r_h = GM_{\rm BH}/v_0$ \citep[where $v_0$ is the velocity dispersion of stars at the center;][for our Galaxy, $r_h \approx 0.5$\,pc]{P72}. 
    Following the
    same analysis presented here to set an example, we estimate 
    that the gravitational energy contribution ($W_{\rm BH}$) that the 
    Milky Way's central black hole excerts on a 3\,pc diameter cloud with
    density $n=10^3$\,cm$^{-3}$, located 5pc
    away from the center of the Galaxy is given by
    $W_{\rm BH}/E_g < 0.01$, which is not significant in comparison
    to the dark matter contribution.

    On the other hand, the local dark matter background permeating a
    molecular cloud is shown to be able to contribute to its star
    formation if the appropriate conditions are fulfilled.  The high dark
    matter densities and low velocity dispersions, found in dwarf
    spheroidal galaxies, needed for the tidal energy to be important for
    the evolution of molecular clouds are not expected in galaxies where
    there is a significant amount of molecular gas; namely, spiral and
    irregular galaxies. Yet, small galaxies dominated by dark matter with
    a significant fraction of gas are not ruled out. Moreover, although
    the Solar Neighborhood does not meet the dark matter conditions where
    star formation would be greatly enhanced by its presence, we predict a
    small contribution for big clouds.

    This work has shown that despite the huge difference between the
    length scales of molecular clouds and dark matter halos, the latter
    could influence the evolution of a star forming region.  We show that
    this effect is negligible in our local environment, but that it can
    have possible applications in other environments, other locations or
    other galaxies.  In summay, our study firmly suggests that the
    external dark matter gravitational contribution should be considered,
    when analyzing the dynamical evolution of a molecular cloud, if the
    right conditions are fullfilled.

    \section*{ACKNOWLEDGMENTS}
    We are grateful to A. Kravtsov for the numerical code used in our study.
    This work has received partial support from grants UNAM/DGAPA IN110409 to JBP.
    ASM acknowledges CONACyT master's degree grant.
    This work makes extensive use of the NASA-ADS database system.
    We thank an anonymus referee for a careful reading of the manuscript.


    \end{document}